\documentclass[final]{svjour2}
\usepackage{graphicx}
\usepackage{rotating}
\usepackage{amssymb}
\usepackage{mathptmx}
\usepackage[numbers]{natbib}
\usepackage{amsmath}
\usepackage{upgreek}
\makeatletter
\journalname{Journal of Low Temperature Physics}

\bibpunct{}{}{,}{s}{}{,}

\begin{document}

\newcommand{\hdblarrow}{H\makebox[0.9ex][l]{$\downdownarrows$}-}
\title{Transition-Edge Sensors for Particle Induced X-ray Emission Measurements}

\author{M. R. J. Palosaari$^1$ \and K. M. Kinnunen$^1$  \and J. Julin$^2$ \and M. Laitinen$^2$ \and M. Napari$^2$ \and T. Sajavaara$^2$ \and W. B. Doriese$^3$ \and J. Fowler$^3$ \and C. Reintsema$^3$ \and D. Swetz$^3$ \and D. Schmidt$^3$ \and J. Ullom$^3$ \and I. J. Maasilta$^1$ }

\institute{1: Nanoscience Center, Department of Physics, P. O. Box 35, FI-40014 University of Jyv\"askyl\"a, Finland\\
Tel.:+358 40 8054446
\email{mikko.palosaari@jyu.fi}
\\2: Accelerator Laboratory, Department of Physics, P. O. Box 35, FI-40014 University of Jyv\"askyl\"a, Finland
\\3: National Institute of Standards and Technology, Boulder, Colorado 80305, USA}

\date{12.07.2013}

\maketitle

\begin{abstract}

In this paper we present a new measurement setup, where a transition-edge sensor detector array is used to detect X-rays in particle induced X-ray emission measurements with a 2 MeV proton beam. Transition-edge sensors offer 
orders of magnitude improvement in energy resolution compared to conventional silicon or germanium detectors, making it possible to recognize spectral lines in materials analysis that have previously been impossible to resolve, and to get chemical information from the elements. Our sensors are cooled to the operation temperature ($\sim$ 65~mK) with a cryogen-free adiabatic demagnetization refrigerator, which houses a specially designed X-ray snout that has a vacuum tight window to couple in the radiation. 
For the best pixel, the measured instrumental energy resolution  was 3.06~eV full width at half maximum at 5.9~keV. We discuss the current status of the project, benefits of transition-edge sensors when used in particle induced X-ray emission spectroscopy, and the results from the first measurements.

\end{abstract}

\pagebreak

\section{Introduction}
In particle induced X-ray emission (PIXE) measurements an energetic ion beam, normally a 1--4~MeV hydrogen or helium ion-beam, is directed to the sample to be analyzed, and sample composition for elements typically heavier than Na or Mg can be determined by means of characteristic X-ray emission \cite{johansson}. Over the last four decades PIXE has developed to become a standard tool for elemental analysis in many fields of science. In geology, art restoration and medical diagnostic, to name a few, PIXE has been used to determine the elemental composition of samples \cite{geo,art,med}.

The development of micro-PIXE, where the particle beam is focused to about 1~$\upmu$m diameter spot, has opened the possibility to raster the target in order to get the positional elemental information of the sample. One interesting special feature of PIXE is that it is possible to irradiate samples with ion beams in ambient conditions, thus making the analysis of biological and large specimens, like paintings, possible. In addition, compared to electron beam excitement, the bremsstrahlung from protons is negligible because of their higher mass \cite{johansson}. This means that the continuum X-ray background is orders of magnitude smaller compared to scanning electron microscope energy-dispersive X-ray spectroscopy (SEM-EDX). 

Traditionally, reverse biased silicon or germanium detectors have been the work horse for PIXE because of their wide dynamic range in energy, ease of use and reasonable cost. These properties have so far compensated the limited energy resolution. On the other hand, wavelength dispersive detectors with their great energy resolution have been successfully used with PIXE \cite{wds}, but their intrinsically low throughput makes the measurements cumbersome and time consuming.

By using transition-edge sensor (TES) microcalorimeters operating at cryogenic temperatures, one combines the benefits of energy dispersive detectors (efficiency, wide energy range) and wavelength dispersive detectors (resolution) \cite{moseley}. TES detectors have matured to the state where they are used in number of applications, thanks to their superior energy resolution and sensitivity. A TES is a device that operates between the superconducting and the normal state of a metallic thin film with a coupled absorber \cite{irwin}.

Within the superconducting transition region, the resistance of a TES is very sensitive to changes in temperature.  The device is connected to a heat bath via a weak thermal link so that when a photon hits the absorber and is converted into heat, due to the small heat capacity of the detector, a relatively large temperature excursion and change in the resistance of the detector is produced. The change in the current through the TES during the photon event is read out with a superconducting quantum interference device (SQUID), which acts as a highly sensitive current sensor that can be coupled to the low impedance of the TES \cite{irwin}. Multiplexing is needed, when the number of detectors increases and when the number of measurement wires needs to be minimized in order to limit the thermal load to the cryostat.

Demonstrations of combining a single TES detector pixel with PIXE have been published before \cite{porto}, but the energy resolution achieved ($\sim$ 18~eV at 1.7~keV) has been much worse than in this work, and quite far from the expected theoretical limits of $\sim$ 1.3~eV for a typical TES detector at that energy \cite{nisttes}. 

\section{Experimental Setup}
\begin{figure}[ht]
\begin{center}
\includegraphics[width=0.65\linewidth,keepaspectratio]{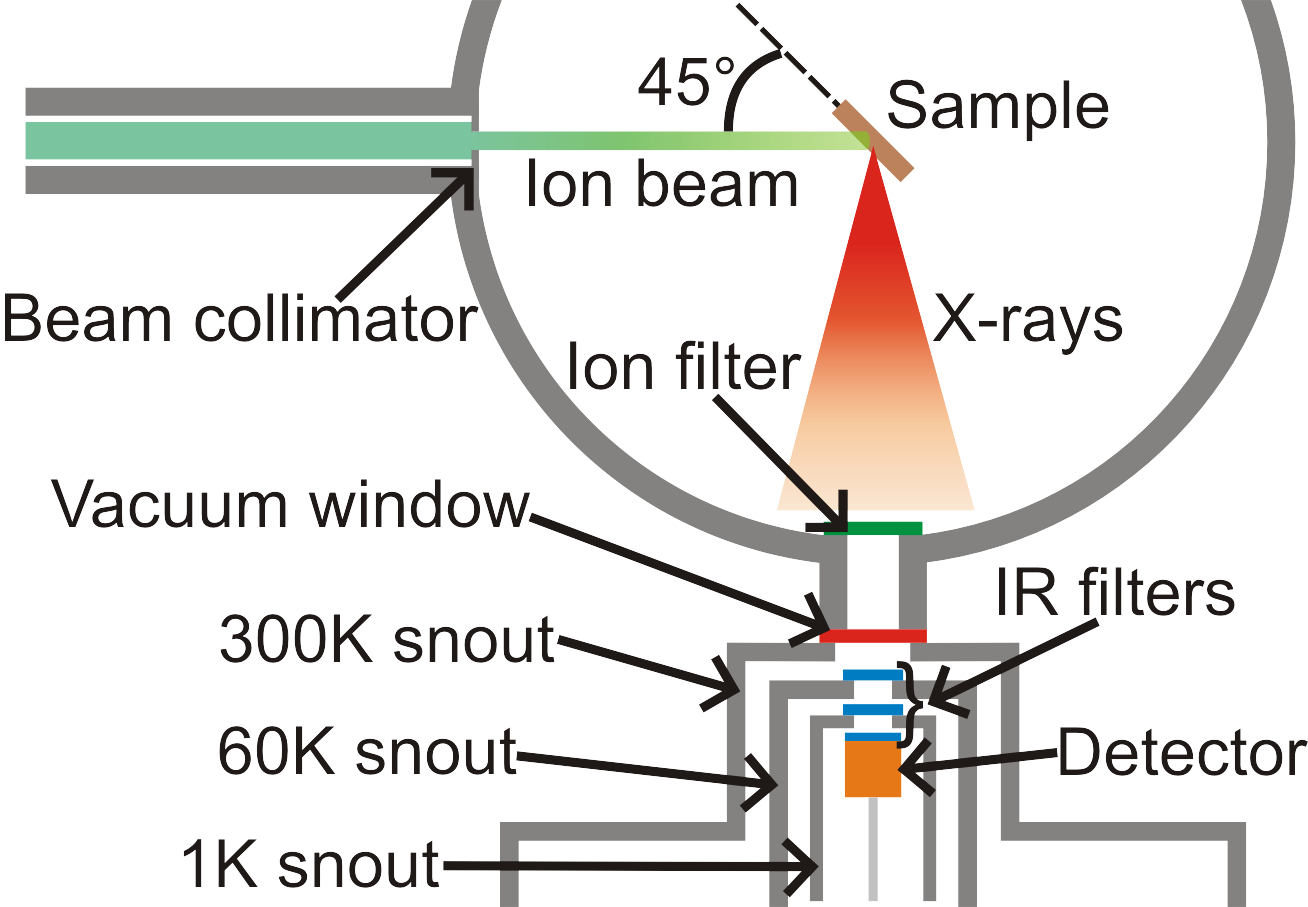}
\end{center}
\caption{Schematic image of the target chamber and the ADR snout separated by a vacuum window. The distance from the sample to the detector was 30 cm. The dimensions are not in scale. The SDD measurements were done in a different target chamber. (Color figure online)}
\label{acc}
\end{figure}

A Pelletron accelerator with 1.7~MV maximum terminal voltage was used to produce 2.015~MeV incident $^1$H$^{+}$ ion beam that was used in the measurements. The TES X-ray detector array was positioned to an angle of 90$^{\circ}$ with respect to the incident ion beam. The sample was
tilted 45$^{\circ}$ towards the detector (see Fig.~\ref{acc}). The backscattered incident ions were stopped  before reaching the detector snout by means of 80~$\upmu$m Polyethylene terephthalate (PET) film inside the sample chamber, which also filtered out low-energy X-rays. An Amptek X-123SDD silicon drift detector (SDD) (130~eV resolution for 5.9~keV X-rays from $^{55}$Fe source) positioned to an angle of 135$^{\circ}$ with respect to beam line was used as a reference detector.

The microcalorimeter TES detector array consisting of 160 pixels was fabricated at NIST Boulder. The superconducting thin film is a molybdenum-copper bilayer film, where the proximity effect \cite{proximity} is used to achieve a critical temperature of $\sim$ 100~mK. The absorber is made of bismuth with horizontal dimensions of 350~$\upmu$m$\times$350~$\upmu$m and with 2.5~$\upmu$m thickness.  A collimator with a 320~$\upmu$m $\times$ 305~$\upmu$m aperture size for each pixel is placed on top of the detector chip to stop photons from hitting the substrate. The separation between the detector and the collimator is 20~$\upmu$m. The X-rays from the sample chamber enter through a vacuum window (AP3.3 ultra-thin polymer/Al silicon grid X-ray window by Moxtek Inc.) into the cryostat.  Inside the detector snout, the X-rays still have to penetrate through three layers of IR radiation filters (each a 1~$\upmu$m PET film with 1~$\upmu$m aluminum coating) at three different temperature stages of the snout (60~K, 1~K, 50~mK), before impinging on the TES detector array. 

In these first measurements only 12 pixels were connected out of which 9 gave good data. One pixel had major problems with SQUID locking and did not produce good enough data and two pixels had over an order of magnitude smaller count rates and their data was removed from the final spectra due to energy calibration problems of insufficient number of events.  The read-out of the detectors was realized with NIST time-division-multiplexing (TDM) SQUID electronics \cite{tdm}.  In TDM, many microcalorimeters are read out in a single set of wires by turning the SQUIDs on sequentially.

The cryogenic cooling was achieved with an adiabatic demagnetization refrigerator (ADR) (Denali Model 102, High Precision Devices, Inc). The cryostat has a special snout designed to couple the X-rays into the detectors, with two layers of cylindrical magnetic shielding from A4K material (Amuneal Corp.) at 60~K and 1~K, and one cylindrical superconducting Al shield at 50~mK. The pre-cooling of the ADR's two stages (60~K and 3~K) is done with Cryomech pulse-tube refrigerator model no. PT407 RM, making the whole system cryogen-free. Vibrations from the pulse tube are minimized by using a remote valve system and a flexible bellows between the pulse tube and cryostat body. The control software and hardware for the ADR's 4~T magnet was made by STAR Cryoelectronics LLC. The ADR has a base temperature of $\sim$~30 mK and can be regulated at the 65~mK operation point with below 15~$\upmu$K rms accuracy for about 12 hours with the read-out electronics on.  

In this first proof of principle measurement, the detector array was in a location with a solid angle of only $\upOmega$=1.3$\times 10^{-5}$ of 4$\pi$~sr  (for the 12 pixels and only the absorber area visible through the collimator) at a distance of 30~cm from the sample. The solid angle will be increased by a factor of 36 in the future by moving the detector array closer to the sample from 30~cm to 5~cm. Also, by connecting all of the 160 pixels the solid angle will increase by another factor of 13, leading to a total increase by a factor 468.  This increase will translate straight to increase in the X-ray count rates and it gives the possibility to study thin film samples in contrast to the bulk samples discussed here.

\section{Results}
A spectrum of Mn K$\upalpha$ lines  produced by a $^{55}$Fe-source was measured. After the subtraction of the natural line shape and width of the emission, the instrumental energy resolution of the best pixel was 3.06~eV full width at half maximum (FWHM), which is within a few percent of the expected resolution based on the measured average signal and noise. The arithmetic mean resolution of 9 pixels was (3.8 $\pm$ 0.6)~eV each having approximately the same number of events. No external magnetic field optimization was done during the measurements for any of the results.

\begin{figure}[ht]
\begin{center}
\includegraphics[width=0.9\linewidth,keepaspectratio]{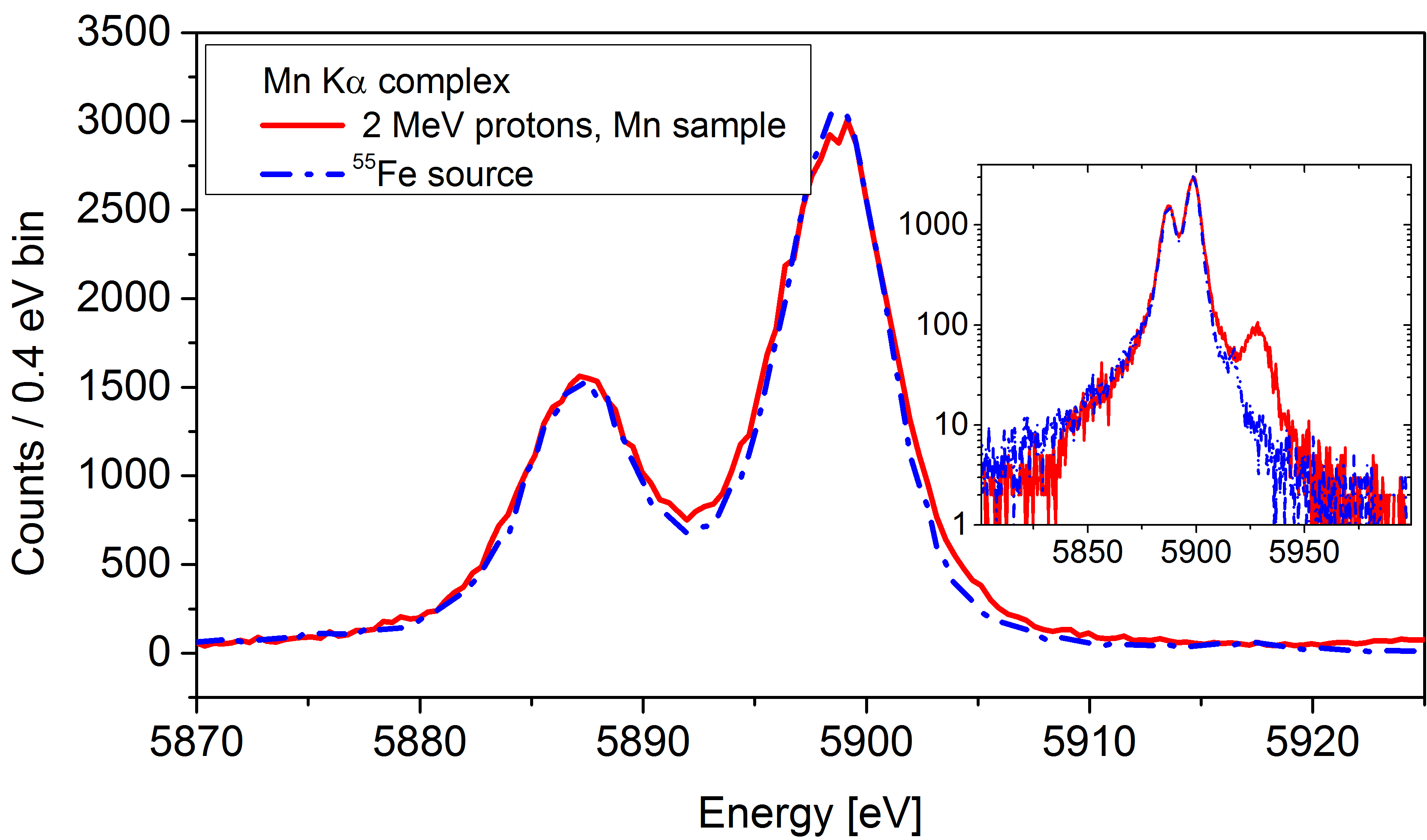}
\end{center}
\caption{ Comparison of Mn K$\upalpha$ complex spectrum of PIXE and $^{55}$Fe source. Spectra are combined from the 9 pixels. Inset shows a satellite peak in the PIXE spectrum discussed later. (Color figure online)}
\label{mnfe}
\end{figure}

In Fig.~\ref{mnfe}, we present the comparison  between the $^{55}$Fe produced Mn spectrum and spectrum from a PIXE measurement with 2~MeV protons and a Mn sample, with the combined data from the 9 pixels. The FWHM broadens by less than 1~eV with the PIXE data compared to the $^{55}$Fe measurement (4.45~eV vs. 3.75~eV respectively). The origin of this broadening is not yet fully understood, but there was some electrical interference present during the Mn PIXE measurement unlike in the $^{55}$Fe measurement, that could affect the energy resolution.

\begin{figure}[ht]
\begin{center}
\includegraphics[width=0.9\linewidth,keepaspectratio]{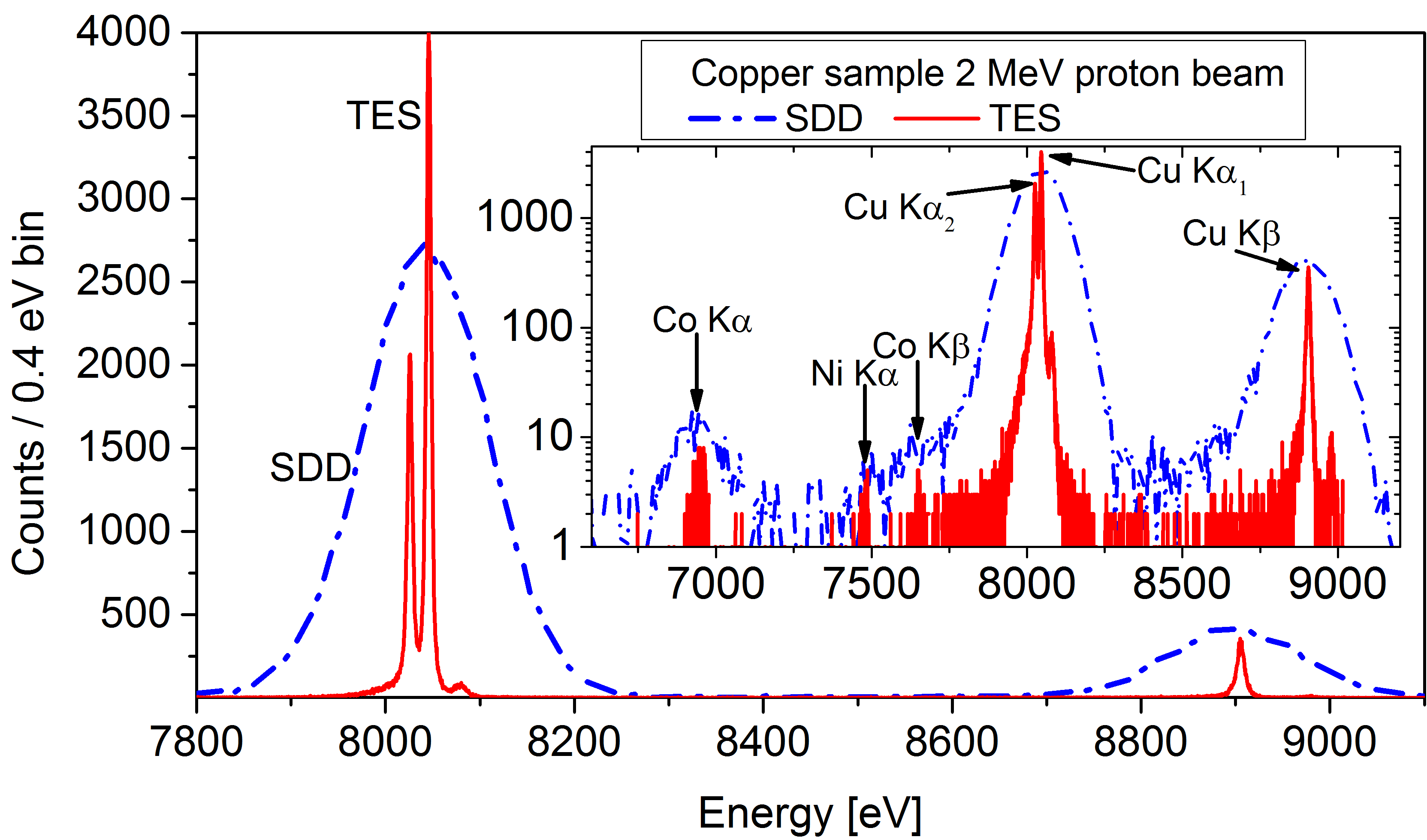}
\end{center}
\caption{Comparison of Cu-sample spectra excited with 2~MeV protons measured with TES (9 pixels) and SDD detectors. Inset shows the same data in logarithmic scale, where the Co K$\upalpha$, K$\upbeta$ and Ni K$\upalpha$ impurity lines of the sample are more visible. The theoretical energy values \cite{data} for the impurity lines are represented with the vertical arrows. (Color figure online)}
\label{Cu}
\end{figure}

A bulk copper sample was also measured with 2~MeV proton excitation, using both TES and SDD detectors to compare the two detector types, with  the results shown in Fig.~\ref{Cu}. The Ni impurities in the sample cannot easily be resolved with the SDD detector but they are clear in the TES data. The satellite peak on the high energy side of the Cu K$\upalpha_1$ and Cu k$\upbeta$ emission line visible in the TES spectrum  is possibly a multivacancy satellite line, where two (or more) inner shell electron vacancies are created simultaneously \cite{sat}. The same feature is also seen in the Mn spectrum in Fig.~\ref{mnfe}.

\section{Conclusions}
In this paper, we have demonstrated Particle Induced X-ray Emission (PIXE) spectroscopy using a multiplexed microcalorimeter TES detector array with an unprecedented energy resolution for PIXE applications. In the future, the beam line and the target chamber will be optimized to maximize the X-ray flux for thin film analysis. This will make it possible to measure not only the elemental composition of thin films in greater detail, but also the chemical environment of the different elements in cases where the chemical shifts are large. Another very interesting future application is the study of the formation of embedded metallic clusters in insulators after implantation and annealing. 
 
\begin{acknowledgements}
This work was supported by the Finnish Funding Agency for Technology and Innovation TEKES, Academy of Finland Project no. 260880 and Academy of Finland Center of Excellence in Nuclear and Accelerator Based Physics (ref. 251353). M. Palosaari would personally like to thank the National Graduate School in Materials Science for funding. 
\end{acknowledgements}

\end{document}